\documentclass[aps,prb,twocolumn,groupedaddress,floatfix,showpacs]{revtex4-2}

\usepackage[T1]{fontenc}

\usepackage{graphicx}
\usepackage{amsmath}
\usepackage{subcaption}
\usepackage{color}
\definecolor{green}{rgb}{0,0.5,0}

\newcommand{\parl}{\parallel}
\newcommand{\beq}{\begin{equation}}
\newcommand{\bearr}{\begin{eqnarray}}
\newcommand{\eeq}{\end{equation}}
\newcommand{\earr}{\end{eqnarray}}

\begin{document}


\title{Rydberg excitons in core-shell nanostructures. \\
II. Plasmon-exciton interaction and enhancement.}


\author{David Ziemkiewicz}
\email{david.ziemkiewicz@utp.edu.pl}
\author{Gerard Czajkowski}
\author{Sylwia Zieli\'{n}ska-Raczy\'{n}ska}
 \affiliation{Department of
 Physics, Technical University of Bydgoszcz,
\\ Al. Prof. S. Kaliskiego 7, 85-789 Bydgoszcz, Poland}


\date{\today}




\begin{abstract}
Interaction between Rydberg excitons in core-shell Cu$_2$O nanostructures and  surface plasmons is investigated. An enhancement of a quadrupole transitions to the Rydberg exciton states in copper oxide plasmonic nanaostructures is examined. A  possibility of a modification of plasmonic enhancment (i.e., an average amplification factor) of electric field distributions  in the vicinity of nanostructures, which depends on its shape and geometrical size, is also discussed. \end{abstract}



\maketitle

\section{Introduction}

An interaction of light with nanoscale matter remains one of the most dynamic and significant frontiers in modern optics and nanophotonics. Excitons in semiconductor nanocrystals and plasmons in metal nanosurfaces are the examples of quasiparticles which govern optical processes in nanostructures.  Both  stand out for their exceptional tunability, strong light-matter interactions and relevance for  both fundamental science and modern technologies.

As it was mentioned in our previous paper I \cite{Paper1}, Rydberg excitons characterized by a particularly large principal quantum number $n$ first observed in Cu$_2$O \cite{Kazimierczuk}, are the bound  electron-hole pairs  with  giant microscopic dimension $\sim n^2$ (up to 1$\mu$m for $n$=25 in Cu$_2$O) and size-dependent transition frequencies $\sim n^{-3}$.  The significant  dipole moments resulting from a large dimension lead to giant exciton dipole-dipole interaction. Another unique feature of Cu$_2$O excitons is their unusual stability which is due to the particularly large binding energy of 90 meV and the lifetimes $\sim n^3$ reaching nanoseconds \cite{scaling}.

Plasmons arise from the collective oscillations of free electrons in metals. When the dimensions of a metal particle are reduced to the nanoscale and become comparable to or smaller than the wavelength of the incident light, these oscillations become spatially confined, giving rise to localized surface plasmons. Surface plasmons polaritons (SPPs) are electromagnetic excitations formed on an interface between two media characterized by opposite signs of the dielectric permittivity (typically metal and dielectric) at some particular frequency. They propagate along metal/dielectric interface. 
Unlike electromagnetic waves in free space, the plasmon modes are bound to the material interface,
so  they can be guided by the structure \cite{Barnes}, leading to miniaturized photonic circuits. The group velocities of propagating SPPs modes strongly depend on the optical properties of surrounding materials, which gives a possibility to control their dynamics and possibly their interaction with any objects in the vicinity \cite{my2018}. 
 
A second important characteristic of surface plasmons is, that compared to waves propagating in free space, SPPs are usually characterized by smaller wavelength and higher field intensity \cite{Stauber}, forming strongly localized, subwavelength field maxima. Crucially, the spatial distribution of the plasmon’s electric field can be manipulated with an appropriate choice of the geometry of the structure sustaining plasmons (typically referred to as nanoantennas). This capability to tailor the field distribution means that the structure can be coupled to the EM field of an atom, a molecule, a quantum dot or an exciton placed in vicinity, altering its radiative and nonradiative properties \cite{gian}. The amplification of various emission processes through a metallic nanoantenna has been demonstrated on a single-molecule level \cite{Kink} and more recently, investigations have focused on plasmon-exciton systems \cite{Zhou}. A coupling  light to plasmon resonances in metal nanostructures has attracted attention primarily because of plasmonic ability  to concentrate optical fields to volumes well below the diffraction limit, enhanced optical absorption and scattering at characteristic resonance frequencies determined by the metal dielectric properties, particle size, shape and surrounding medium \cite{gogoi,Valagio}.

A new phenomena can arise when plasmons and excitons in nanostructures are coupled one to  another \cite{gogoi} and extending a study to Rydberg exciton-plasmon interaction is a natural  step forward. Moreover, what makes this topic exceptional   and worth studying is the fact that both  Rydberg excitons and plasmons are large enough to exist on the border between classical and quantum realm.
Rydberg excitons are unique quantum objects capable of strong,  linear and nonlinear interactions;  
their large size makes them particularly well suited for experiments with micrometer-size plasmonic systems. It should be stressed that Rydberg exciton lifetimes are comparable to typical plasmonic lifetimes, which enables  exciton-plasmon coupling feasible and their interactions particularly interesting, yet unexplored  area of research. Specifically, typically achieved plasmon lifetimes are about $10^{-14}$-$10^{-10}$ s \cite{berini}, compared to exciton lifetime from $10^{-12}$ s to $10^{-10}$ s. Similarly, the range of exciton radius which extends from nanometer ($n$=1) to  micrometers for $n>$25 \cite{Kazimierczuk} matches the typical propagation distance of plasmons. Therefore, fabrication of plasmonic structures capable of interacting with confining and guiding single excitons may be feasible. 

The plurality of Rydberg excitonic states in copper oxide offers a wealth of available quantum one- or two-photon excitations in a wide range of frequency, from optical to microwave \cite{Heck017, Kim, konwerter, maser}. Apart from the strong dipole-allowed transitions from the ground state to $P$ excitonic states, giving rise to the characteristic excitonic absorption spectrum, there are other, weaker quadrupole transitions  to $S$ or $D$ states \cite{Chakra, Sylwiaqb}. The general concept of  enhancement of quadrupole transitions with plasmons based on the fact that plasmonic structures are capable  of focusing light into very small spots in which highly focused field is characterized by a particularly large electric field gradient; this in turn affects quadrupole transitions.

To observe and moderate the new effects in plasmon-exciton coupling, it is necessary to apply nanoscale systems with plasmonic and excitonic components whose size, shape and other physical properties are controlled. Therefore the idea of studying core-shell nanostructures with a central metallic core wrapped in a distinct outer shell with excitons seems to be worth examining in the context of steering plasmon-exciton interaction. Such a system provides a possibility of spectral tunability: by adjusting  the core size or shell thickness one can shift the resonance frequency \cite{Valagio}, the plasmon modes can be split into lower and higher energy bonding states  by  an interaction between the inner core surface and an outer shell, while a local electric field can be tuned through changes in geometry and material composition \cite{Wang}. The first attempts of considering plasmon-exciton interactions in a hybrid core-shell system (silver/organic) were made by several groups \cite{Antos2014, Stete, Petoukhoff, Kondorsky}, who discussed optical spectra and plasmon-exciton coupling in such metal-organic systems. 

This paper is structured as follows. We start by introducing  dispersive properties of plasmons, with a particular emphasis on spherical metal nanoparticles. Then, an enhancement of the quadrupole transitions of Rydberg excitons is discussed. After brief theoretical description, an averaging method is introduced to take the large size of Rydberg excitons into account. Section IV is devoted to the calculation results in three selected geometries. In section V, conclusions are presented. The paper ends with two appendices discussing the calculation of exciton wavefunction and Finite-Difference Time-Domain (FDTD) calculation of the plasmonic field, respectively.
 
\section{Surface plasmons} 
A surface plasmon is a collective excitation of electromagnetic field and free charges in metal, existing on the interface between metal and dielectric. It propagates as a wave characterized by a wavenumber \cite{Chubchev}
\begin{equation}\label{dys_plazm}
k(\omega)=k_0\sqrt{\frac{\epsilon_1\epsilon_2}{\epsilon_1+\epsilon_2}},
\end{equation}
where $\epsilon_1(\omega)<0$ is the permittivity of metal, $\epsilon_2(\omega)>0$ characterizes the dielectric and $k_0=\frac{\omega}{c}$ is the free space wavenumber. Propagating modes are characterized by a real value of $k$. This means that two conditions have to be fulfilled: $\epsilon_1\epsilon_2<0$ and $\epsilon_1+\epsilon_2<0$. Metals typically used in plasmonics such as silver, gold, copper satisfy this condition in the optical frequency range. In particular, similarly to our previous works \cite{copper_plasm}, here we focus on surface plasmons on copper-Cu$_2$O interface. For these two materials, the permittivity $\epsilon_2 \approx 7.5$ and $\epsilon_1 \approx -7.5$ in the frequency range of excitons. This matching means that $k \rightarrow \infty$ in the limit of negligible losses, implying that deep subwavelength focusing of the plasmon field is possible despite the fact that copper is an inferior conductor compared to gold and silver \cite{copper_plasm}. The above mentioned focusing is a key to quadrupole transition amplification.
 
The metallic sphere in the center of a core-shell particle allows for excitation of dipole and multipole surface plasmon modes \cite{Kelly2003}. In contrast to nanoslits examined in our previous work \cite{my_quad} as well as earlier studies \cite{Okuda2006,Deguchi2009}, a core-shell structure allows for an interaction between surface plasmons and highly confined excitons. In particular, a sufficiently small structure may support only a single exciton, which is a key to single photon applications \cite{Khazali}.    

\begin{figure}[ht!]
\centering
\includegraphics[width=.9\linewidth]{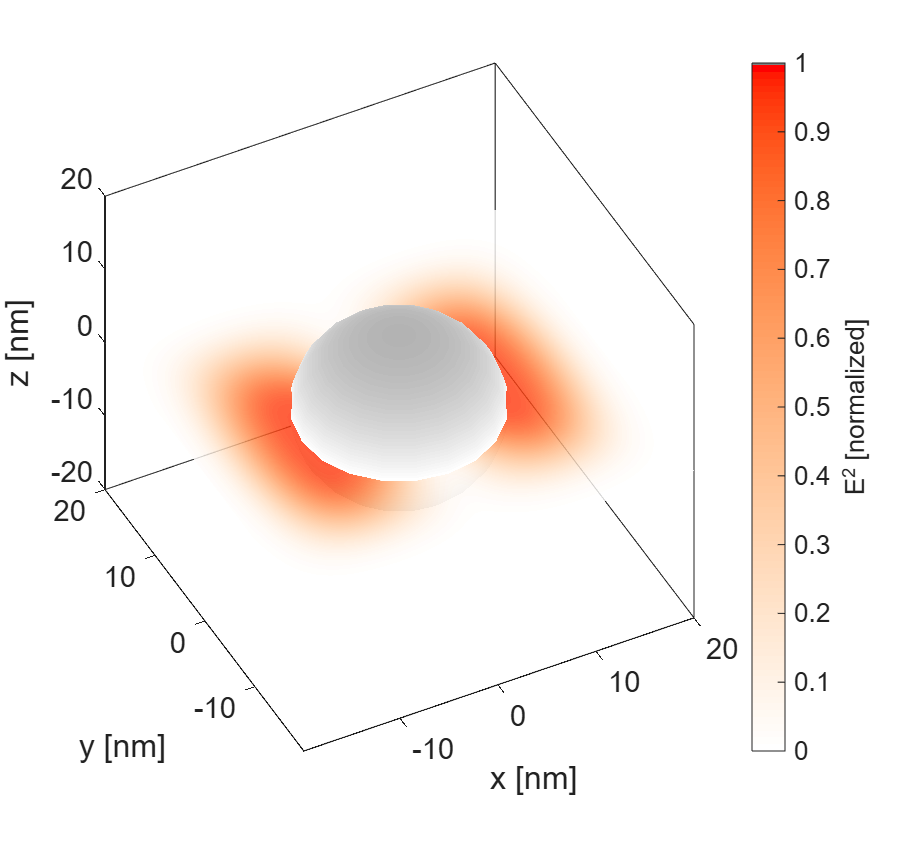}
\caption{Numerically calculated electric field intensity in the vicinity of a metal sphere.}\label{rys_metalball}
\end{figure}

As mentioned above, a surface plasmon is a collective oscillation of the free electrons and associated electric field. In the case of a small metallic sphere, the most fundamental plasmonic mode is a dipole oscillation of the displaced electron cloud \cite{Kelly2003}. In response to the applied field, the conduction electron plasma in the metal forms local concentrations (surface charges) that provide a restoring force to the electrons \cite{Myros2008}. This is illustrated in Fig. \ref{rys_metalball}, where two local maxima of field intensity can be seen near the poles of the nanosphere along the x axis.
In this paper, we use FDTD method to obtain the spatial field distribution of the plasmon in a core-shell structure (see Appendix A for details). This method is commonly used for complicated nanostructure shapes \cite{Myros2008}. 

Since the plasmon field is determined mostly by the properties of the metallic core, it is instructive to study a simpler case of metal sphere without the surrounding shell. In such a system, an analytical solution is possible \cite{Kelly2003}; the electric field of the plasmon is given by
\begin{equation}
E=E_0\hat{x}-\alpha_P E_0\left[\frac{\hat{x}}{r^3}-\frac{3\hat{x}}{r^5}(x\hat{x}+y\hat{y}+z\hat{z})\right],
\end{equation}
where $\hat{x},\hat{y},\hat{z}$ are unit vectors, $E_0$ is the incident electric field and $\alpha_P$ is the polarizability
\begin{equation}
\alpha_P=R^3\frac{\epsilon_1-\epsilon_2}{\epsilon_1+2\epsilon_2},
\end{equation}
where $\epsilon_1$ is the dielectric permittivity of metal, $R$ is the radius of the metal sphere and $\epsilon_2$ is the permittivity of the surrounding medium. The results of FDTD calculations on Fig. \ref{rys_metalball} are consistent with the above solution.

The studies of optical properties of nanospheres in the far field regime are focused on absorption and scattering coefficients, which can be obtained through Mie theory \cite{Kelly2003}. In contrast, here we focus on the near field and its impact on the Rydberg excitons confined in a shell surrounding the sphere. In particular, we study how the local field gradient affects the quadrupole transitions between excitonic states. Thus, our main point of interest is the spatial distribution of the field within the shell.

Apart from the spatial field distribution, it is important to consider the plasmon frequency. Specifically, in order to obtain strong plasmon-exciton interaction, the resonant frequency of the plasmon has to match the exciton frequency. In the case of surface plasmons on bulk surfaces, the frequency of oscillation is determined by the properties of metal (concentration of electrons and their effective mass); for nanoparticles, the frequency is additionally affected by their size \cite{Kelly2003}. This provides a degree of freedom in designing plasmonic structures that are tuned to a specific exciton energy.

Finally, we note that the excitons in Cu$_2$O have a relatively small impact on the dielectric susceptibility, of the order of $10^{-2}$ compared to the bulk value of $\epsilon_2=7.5$. Therefore, one can assume one-way interaction where the excitons are affected by the plasmonic field, but the surface plasmons are not strongly affected by excitons. As a result, in our system there is no strong coupling and splitting of the plasmonic absorption peak demonstrated in \cite{Antos2014}.
 
\section{Quadrupole transition enhancement}
\subsection{Point emitter solution}
As mentioned previously, one of the interesting features of plasmonic structures is their capability to enhance quadrupole-allowed transitions between the ground state and exciton $S$ state, already demonstrated experimentally \cite{Neubauer}. In general, this enhancement is a result of substantial electric field gradients in the vicinity of surface plasmons. To understand this mechanism, let's first assume a point-like exciton placed at $\vec{r}=0$. The electric field surrounding the exciton can be expressed by a Taylor series, with the first two terms
\begin{equation}
\vec{E}=\vec{E}(0)+\vec{r}\cdot(\nabla \vec{E}),
\end{equation}
where $(\nabla \vec{E})=\left[\frac{\partial E_i}{\partial r_j}\right]_{i,j}$ is the Jacobian matrix of $\vec{E}$. The electric potential $\varphi = -\int \vec{E} \cdot d\vec{r}$ can be then expressed as
\begin{eqnarray}
&&-\varphi = \vec{E}(0)\int d\vec{r} + \int\vec{r}\cdot(\nabla \vec{E})\cdot \vec{dr},
\end{eqnarray}
which results in \cite{Barron1973}
\begin{equation}
\varphi = -\vec{E}(0)\cdot \vec{r} - \frac{1}{2}\sum\limits_{i,j}r_ir_j\frac{\partial E_j}{\partial r_i}
\end{equation}
The energy of exciton is then given by
\begin{equation}\label{eq:energy}
W=-\vec{E}(0)\cdot \vec{p} - \frac{1}{6}\sum\limits_{i,j}Q_{ij}\frac{\partial E_j}{\partial r_i}
\end{equation}
with dipole moment $\vec{p}=q\vec{r}$ and quadrupole moment $Q_{ij}=q(3r_ir_j-r^2\delta_{ij})$ \cite{Jackson,Kern2012}.
The above energy of exciton-field interaction is then used to calculate the transition rate $\gamma$, following the Fermi's golden rule
\begin{equation}\label{eq:gamma}
\gamma \sim \left|\langle\Psi_2|W|\Psi_1\rangle\right|^2,
\end{equation}
where $\Psi_1$ is the initial state of the system and $\Psi_2$ is the final state. In the system considered here, $\Psi_2$ is the excitonic wavefunction, which can be calculated numerically (see Paper I \cite{Paper1}) or, for the case of S excitons, analytically (see Appendix A). The most important aspect of the above estimation is that $\gamma$ is proportional to $(\nabla E)^2$. Therefore, it is beneficial to maximize the electric field gradient in the vicinity of exciton; plasmonic systems provide such amplification by squeezing the field into a subwavelength volume, resulting in very strong local field gradients.

\subsection{Larger excitons}
The basic assumption underlying the above results is that the exciton is point-like, with a negligible size. This approach is valid for many systems studied in the literature, for example Rydberg atoms interacting with 50 nm nanoantenna \cite{Kern2012} and an interaction between nanoslits and 1$S$ exciton in Cu$_2$O \cite{Okuda2006}. In contrast, the recent experimental results by Neubauer et al \cite{Neubauer} involve larger $n \sim 6$ excitons. It has already been  demonstrated in \cite{Takahata2018} that the optical properties of a thin Cu$_2$O layer are affected by nonlocal effects caused by an interplay between the exciton radius and layer thickness. Therefore, in our recent paper \cite{my_quad}, we have proposed an averaging scheme that provides more accurate predictions for the Rydberg states. The following procedure is used
\begin{itemize}
\item The field gradients are averaged over a volume of the exciton, e.g. $(\nabla E)^2$ is replaced by $\hat{(\nabla E)^2}=\langle\Psi|(\nabla E)^2|\Psi\rangle$, where $\Psi$ is the exciton wavefunction.
\item The local amplification factor $\eta(\vec{r})=\gamma/\gamma_0$ is averaged over the shell volume, with an exciton probability density $\rho_{exc}(\vec{r})$. 
\end{itemize}
The final amplification factor integrated over the shell volume is given by
\begin{equation}\label{eq:eta}
\tilde{\eta}=\frac{\int\rho_{exc}(\vec{r})\eta(\vec{r})d^3r}{\int\rho_{exc}(\vec{r})d^3r},
\end{equation}
It should be noted that the infinite potential barrier between core and shell yields a vanishing exciton wavefunction $\Psi(r \rightarrow r_1)=0$. Likewise, the exciton density also approaches 0 at the interface. At the same time, the field gradient is  strongest at the interface. This interplay is discussed in later sections devoted to individual nanostructures.

\subsection{Rydberg blockade considerations}
One of the most outstanding features of Rydberg excitons is the so-called Rydberg blockade. Due to the strong exciton-exciton interactions and the resulting energy shift, an exciton cannot be created in the immediate vicinity of an existing one. In the context of small nanostructures, this means that it is feasible to design systems that can support only one exciton at any given time; this mechanism has been proposed as a way to create a single photon source \cite{Khazali}.

In this paper, we consider core-shell structures with a diameter of the order of $10-10^2$ nm, matching the typical spatial extent of the surface plasmons. The exciton blockade volume can be estimated as \cite{Kazimierczuk}
\begin{equation}
V_{bl} \approx 3\cdot 10^{-7} n^7~\mu m^3,
\end{equation}
where $n$ is the principal quantum number. Therefore, the blockade radius is $r_{bl} = V_{bl}^{1/3} \sim 6.7 n^{7/3}$ nm. Even for $n=5$, the blockade radius is already almost 300 nm, easily exceeding the size of the nanostructure. The possibility to use relatively low principal quantum number $n$ is important considering that the oscillator strength of the transition quickly decreases with $n$ (for $P$ excitons, it follows $n^{-3}$ scaling \cite{Kazimierczuk}). Furthermore, if the goal is to maximize the power emitted/absorbed by the nanostructure, it becomes beneficial to maximize the number of excitons inside it. For a given structure volume $V$, the number of excitons will be $N<V/V_{bl}$, so that $N \sim n^{-7}$, further increasing the importance of using relatively low principal quantum numbers. On the other hand, if we consider a quadrupole transition between two exciton states such as $nS$ and $nD$, described by wavefunctions $\Psi_1$ and $\Psi_2$ correspondingly, then the transition rate given by Eq. (\ref{eq:gamma}) will scale as $n^4$ due to the factor $r^2$ in the quadrupole transition moment $Q_{ij}$. Therefore, in this case it is beneficial to use highly excited Rydberg states. Such a scenario is highly relevant to the recent studies of quantum beats \cite{Thomas,myQB}.

\section{Calculation results}
We have examined the enhancement of the quadrupole transition from the ground state to an excited $S$ exciton state   in a set of nanostructures. Numerically calculated plasmon field distributions are used to estimate the enhancement factor. the spatial averaging described above is used to take account for the exciton size; comparisons between low $n$ excitons and Rydberg states are made.

\subsection{Spherical structure}
\begin{figure}[ht!]
\centering
a)\includegraphics[width=.8\linewidth]{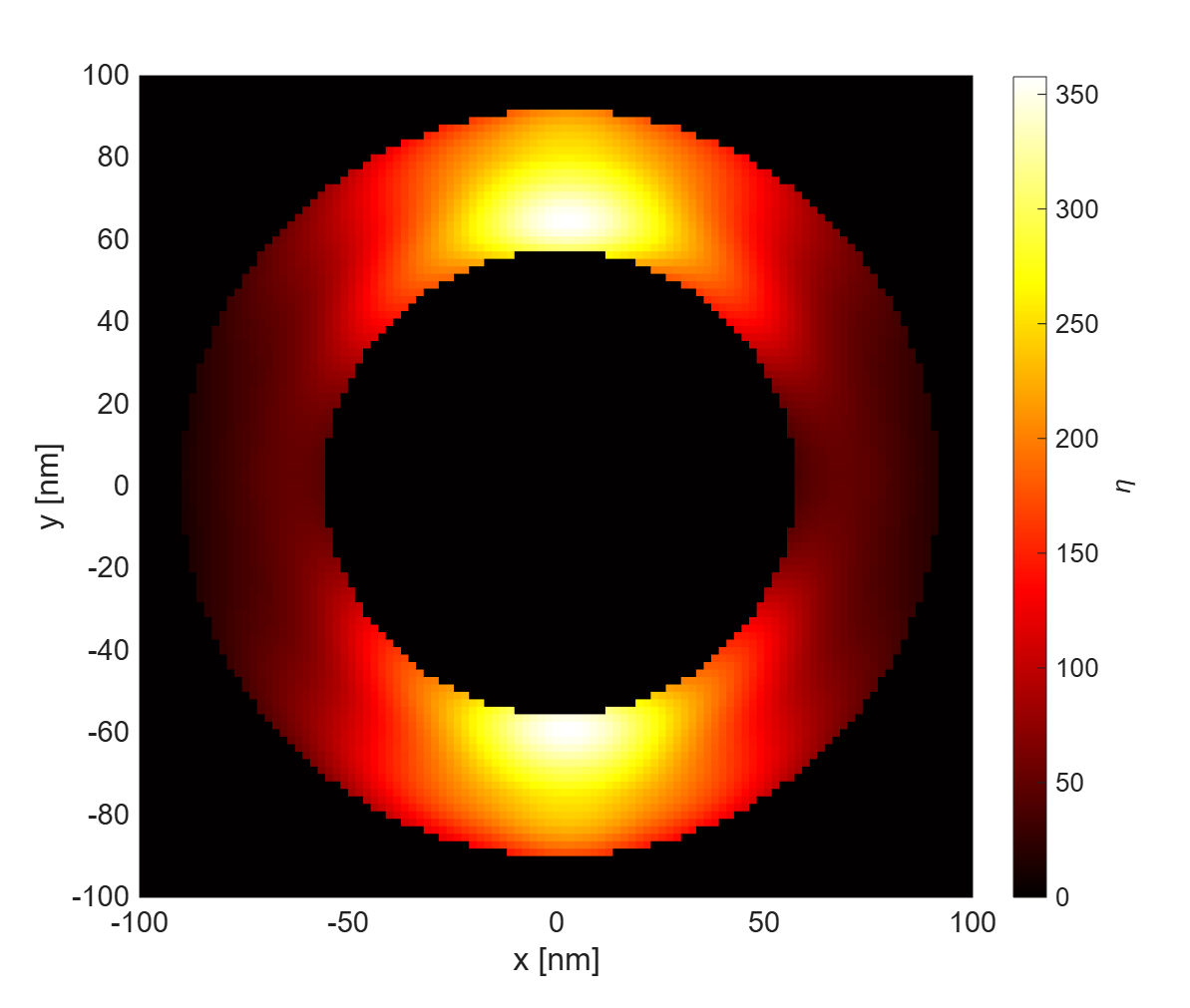}
b)\includegraphics[width=.8\linewidth]{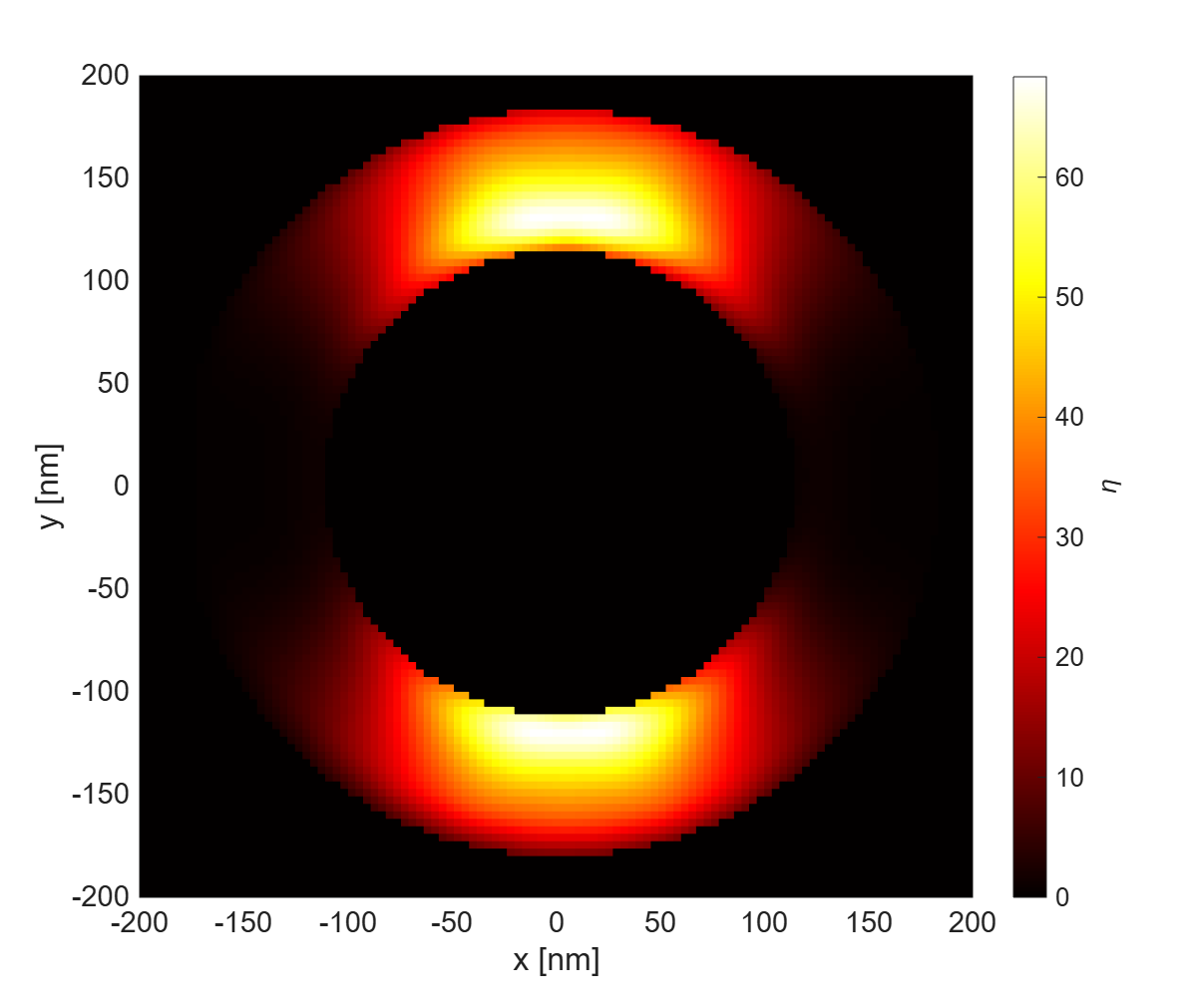}
\caption{Local amplification (color) of the quadrupole transition to 3$S$ state for 100 nm and 200 nm diameter core. Shell thickness is half of the core radius.}\label{rys_1}
\end{figure}

As discussed in the previous section, a spherical core-shell structure supports mostly dipole- and quadrupole plasmonic modes. The spatial distribution of the field of these modes is related to the radius of the metallic core and exhibits deep subwavelength focusing, resulting in a strong field gradient. Due to the rotational symmetry of the plasmonic dipole mode, the two-dimensional cross-section is sufficient to illustrate its field distribution. Such cross-sections are shown in Fig. \ref{rys_1}. Specifically, the amplification of the transition to the $3S$ exciton state is presented.
In the case of the smaller sphere, the peak enhancement calculated from Eq. (\ref{eq:eta}) is of the order of $\eta \sim 350$. There are two factors contributing to this value; in a plasmonic resonance, the local field strength can exceed the incident field strength by a factor of $10^2$ \cite{Kelly2003}. Furthermore, the local field maxima are considerably smaller than the incident field wavelength ($\lambda=570$ nm in vacuum, $\lambda'=210$ nm inside Cu$_2$O), resulting in stronger field gradients. For the larger structure (Fig. \ref{rys_1} b)), the plasmonic mode is less confined, resulting in a lower peak amplification $\eta \sim 70$. However, in the second structure the volume of the Cu$_2$O shell is larger by a factor of 8 and accordingly the number of excitons that can exist inside the shell is also larger by the same factor, so that the total power absorbed by the exciton transition is larger for the larger structure. An interesting feature of the amplification distribution is the fact that it is nonzero even in the areas where the plasmonic field vanishes (here, left and right edge of the circle). This is caused by the finite size of the exciton ($\sim 20$ nm diameter for 3$S$ state); the amplification is averaged over exciton volume and always includes regions with nonzero field gradient. The effect is stronger in the smaller structure, where the exciton size is a larger fraction of the structure size; one can see that in Fig. \ref{rys_1} a) the lowest amplification is of the order of 30 while in Fig. \ref{rys_1} b) it is below 10.

\begin{figure}[ht!]
\centering
\includegraphics[width=.9\linewidth]{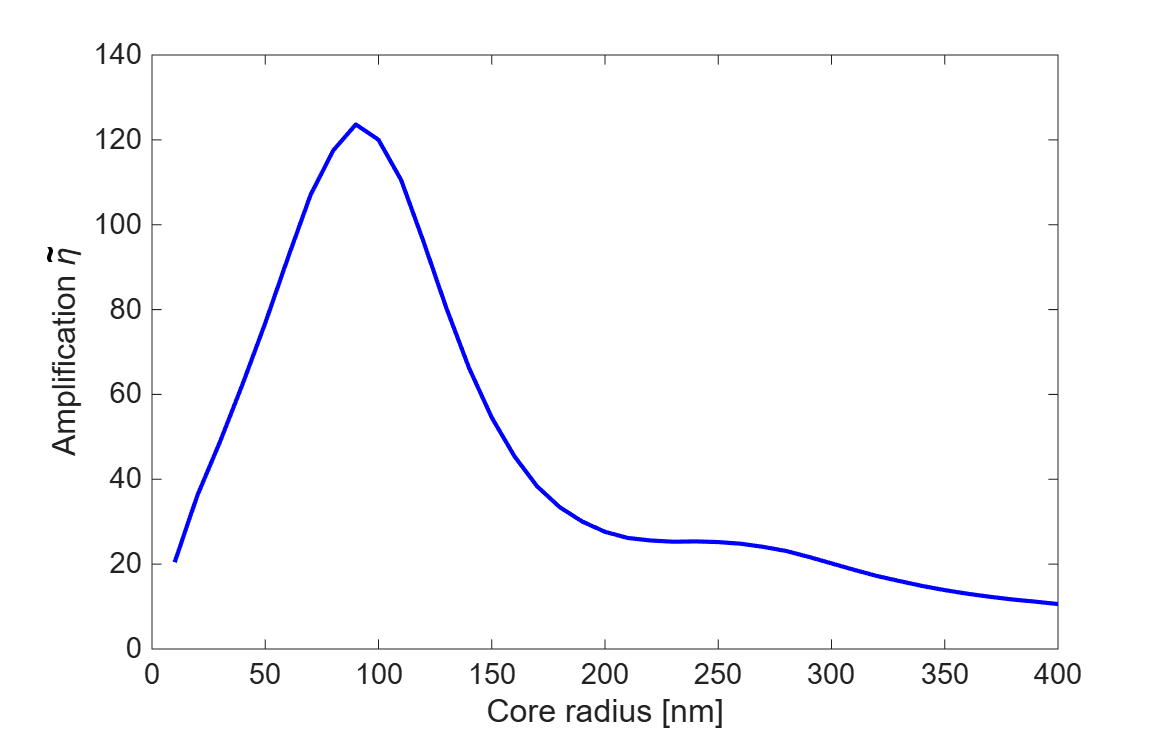}
\caption{Averaged amplification of the quadrupole transition to 3$S$ state in a core-shell structure.}\label{rys_1b}
\end{figure}

The total amplification $\eta$ averaged over the shell volume is presented in Fig. \ref{rys_1b} as a function of the core radius. In all calculations the shell thickness was $h=40$ nm.
The large maximum at $r \approx 80$ nm corresponds to the case where a standing wave pattern can form inside the shell, helping to establish a strong plasmonic resonance. Specifically, the average shell radius is 100 nm and the circumference is approximately 630 nm, which is 3 times larger than the incident field wavelength inside Cu$_2$O. For a smaller core radius, amplification is reduced due to the fact that the metal core becomes a small fraction of the entire structure and also its plasmonic field is weaker (core polarizability scales with its radius \cite{Kelly2003}). For very large structures, amplification is reduced due to a small field gradient. Specifically, the field distribution becomes dominated by waveguide-like mode traveling through the shell while subwavelength surface plasmons become negligible. Another local maximum at $r \sim 260$ nm can be seen  in  Fig. \ref{rys_1b}; it is also related to a standing wave pattern in the shell. The fact that the strongest amplification corresponds to $r \sim 100$ nm means that in such an optimized structure, only Rydberg states $n>7$ will be affected by confinement effects, including a squeezing of exciton wavefunction and an increase of oscillator strength. This suggests that the proposed system is particularly well suited for Rydberg excitons due to the matching spatial scale.

\begin{figure}[ht!]
\centering
a)\includegraphics[width=.8\linewidth]{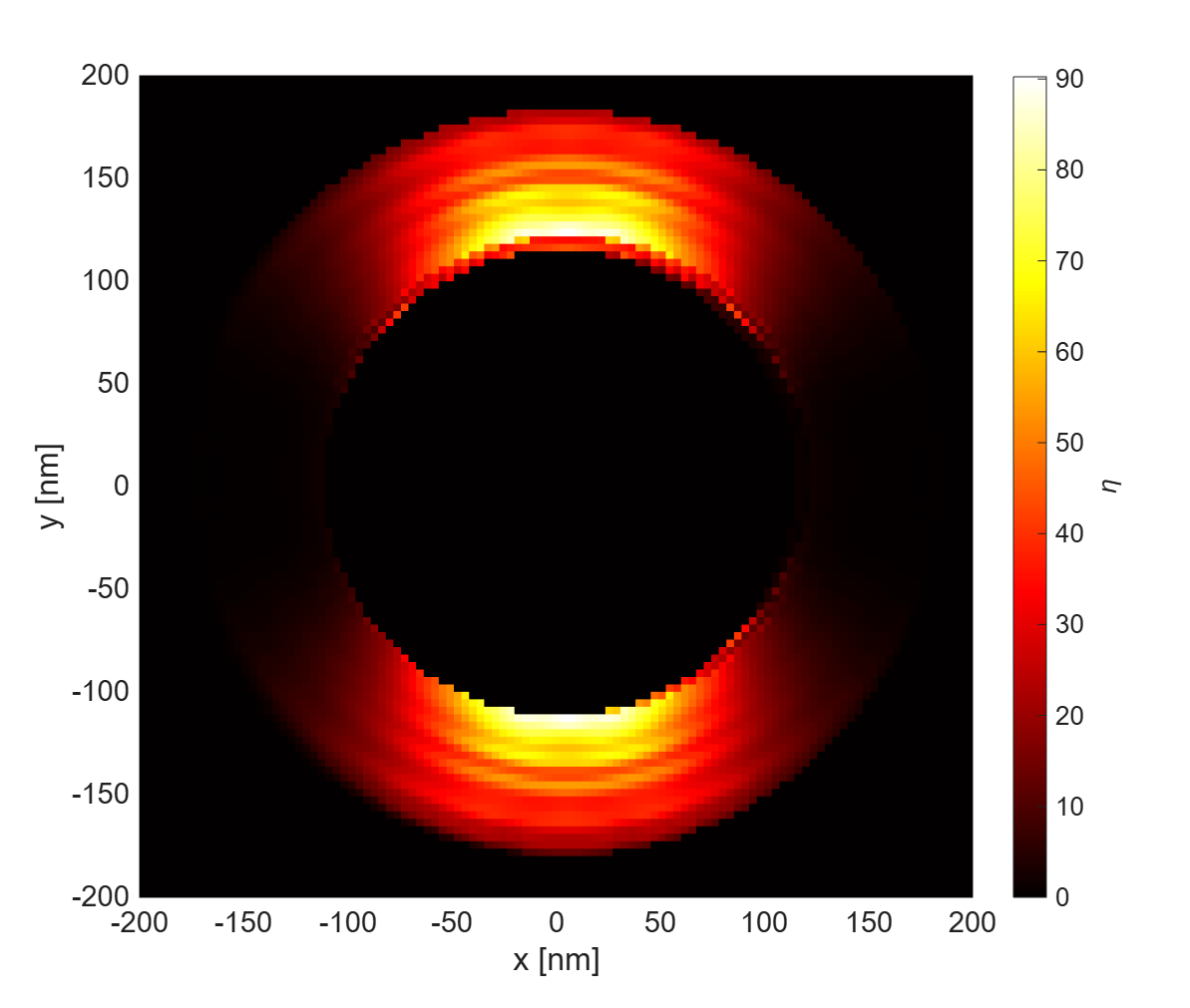}
b)\includegraphics[width=.8\linewidth]{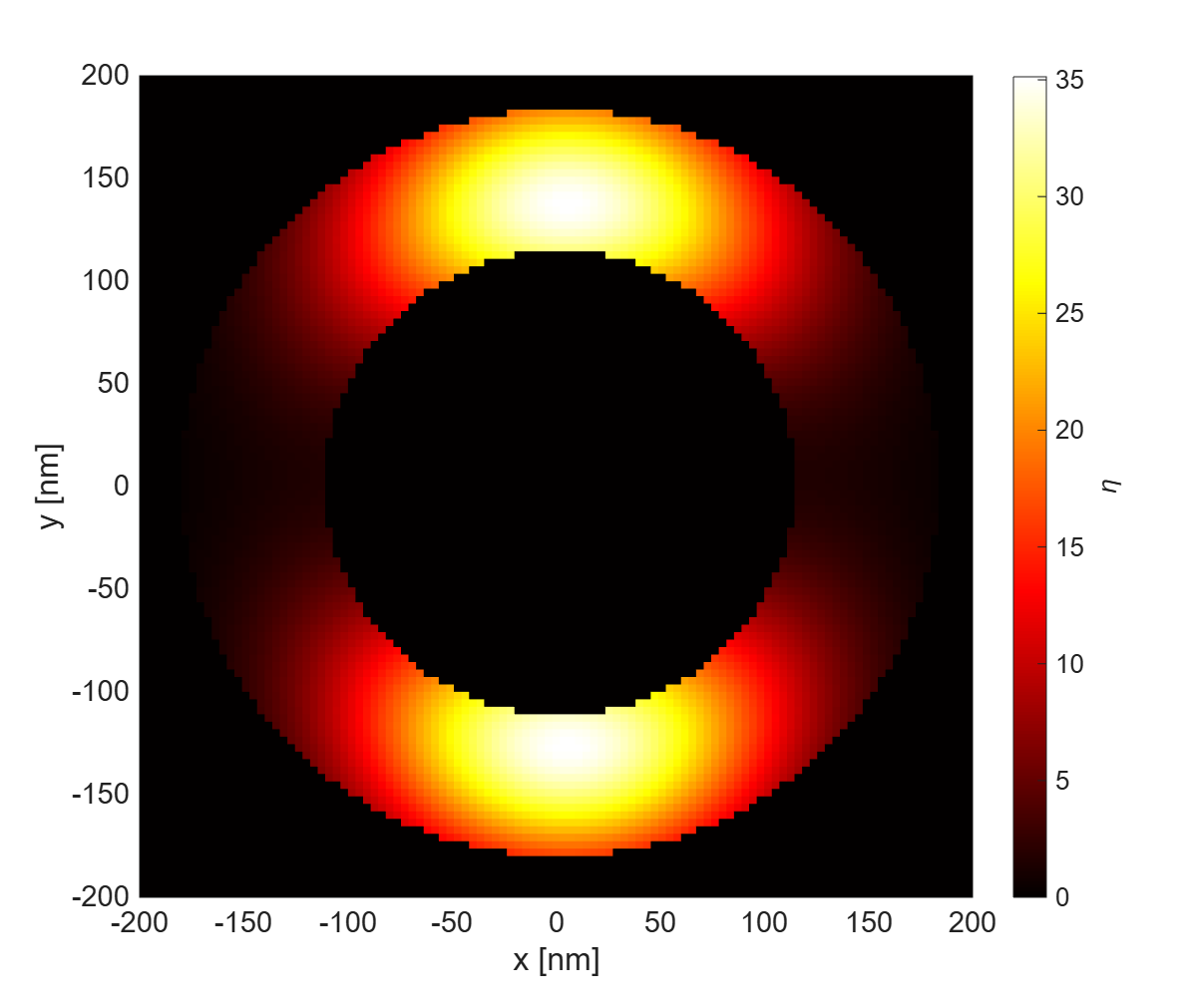}
\caption{Local amplification (color) of the quadrupole transition to a) 1$S$ and b) 6$S$ state for 200 nm diameter core and 50 nm shell.}\label{rys_2}
\end{figure}

To further illustrate the impact of the exciton size, a comparison between $1S$ and $6S$ excitons (1 nm and 36 nm radius, correspondingly) is shown in Fig. \ref{rys_2}.
One can notice several interesting features; the maximum of $1S$ amplification is located close to the metallic surface, where the field gradient is the strongest. As discussed in Paper I \cite{Paper1} and our previous work \cite{my_quad}, the metal surface acts as an effectively infinite potential barrier for the exciton, creating a "dead zone" around the metal where excitons cannot be created. This zone is comparable to the exciton radius and so the small $1S$ exciton can exist very close to the metal surface. On the other hand, the larger $6S$ exciton has to be located further away from the metal layer and so the maximum of amplification is also shifted away from the core (Fig. \ref{rys_2} b)). Due to the stronger field gradient close to the metal surface, the peak amplification is larger for $1S$ exciton. However, that strong amplification is confined in a small volume. As a result, the total  amplification taken as an average over the entire shell is larger for the $6S$ state. Furthermore, the intrinsic quadrupole transition moment of excitons is also larger for higher $n$ states. In conclusion, there is a trade-off between the field confinement and exciton size.

\begin{figure}[ht!]
\centering
a)\includegraphics[width=.8\linewidth]{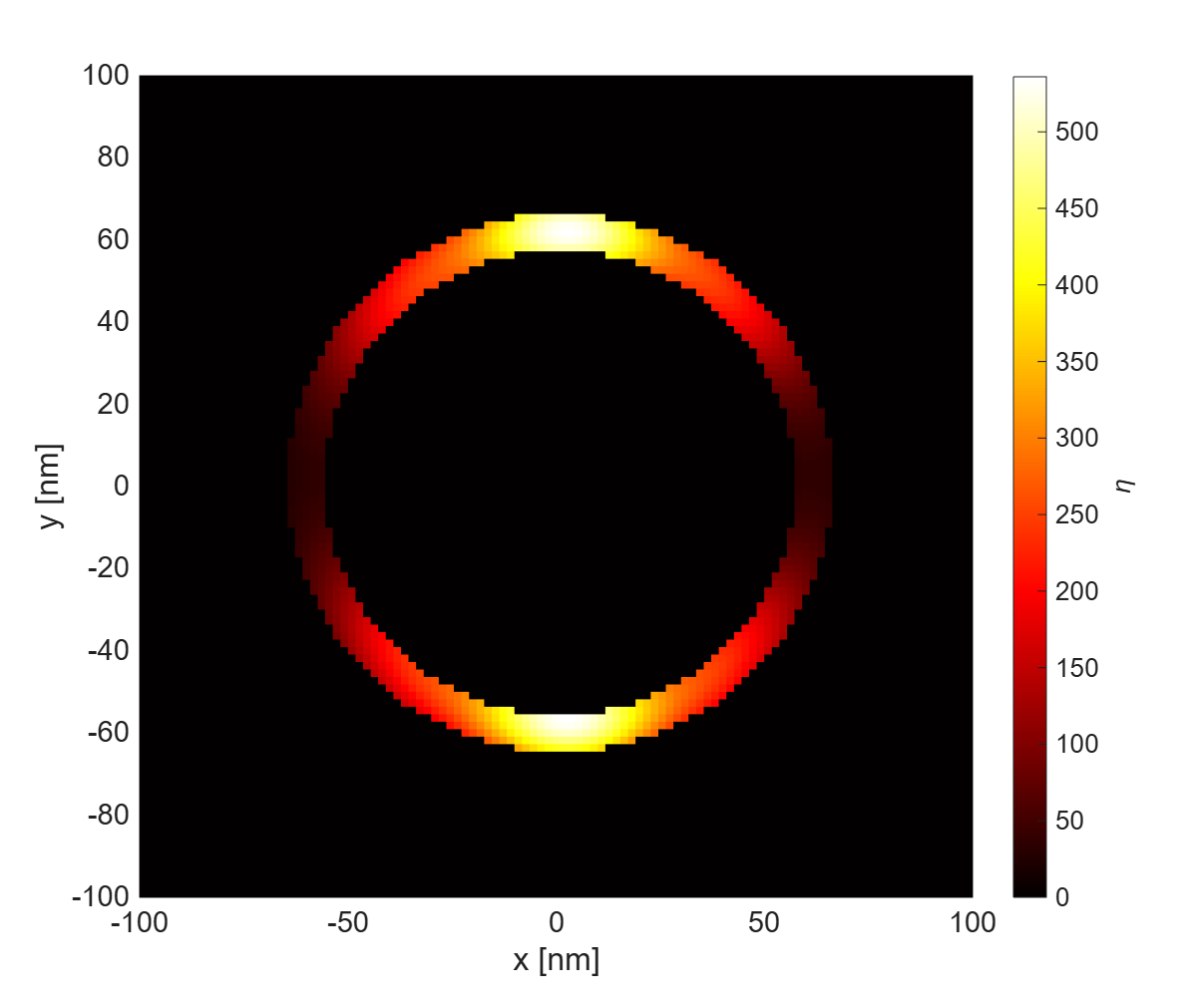}
b)\includegraphics[width=.8\linewidth]{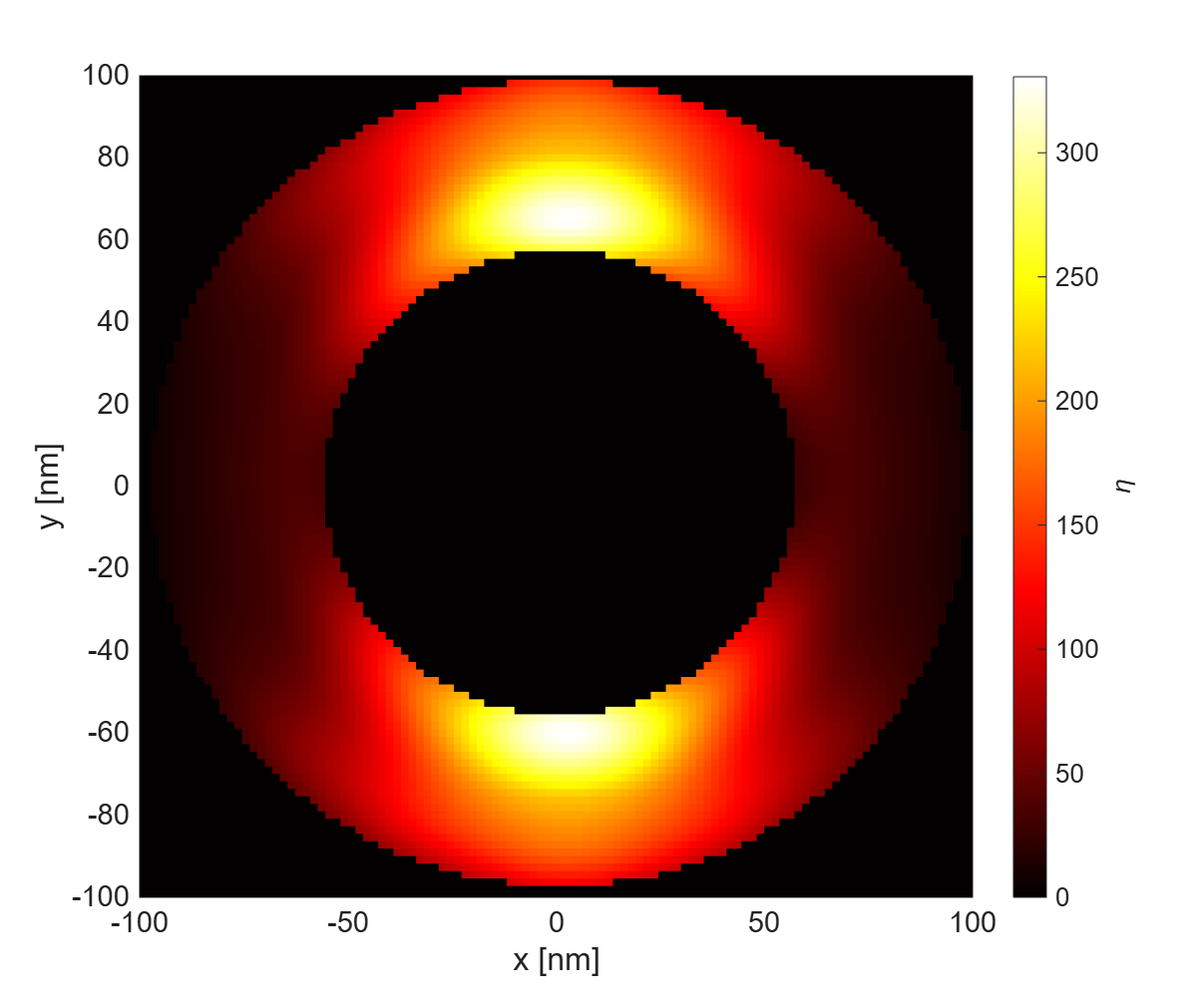}
\caption{Local amplification (color) of the quadrupole transition to 3$S$ state for 100 nm diameter core and a) 10 nm and b) 40 nm shell.}\label{rys_3}
\end{figure}

Finally, one more nanostructure design consideration is the thickness of the Cu$_2$O layer. A comparison between two Cu$_2$O layer sizes is shown in Fig. \ref{rys_3}.
As discussed in \cite{Paper1}, in the case of a thin shell, the exciton state is highly confined; the exciton wavefunction is squeezed in the radial direction but expands in other directions. As a result, the averaging over exciton volume takes into account a larger fraction of the shell. This is reflected in Fig. \ref{rys_3} a), where the lowest amplification is of the order of 120. The peak amplification is over 500 due to the fact that a thin shell helps to further confine the plasmonic field. In a thicker layer, the peak amplification is reduced to $\eta \sim 350$ and the minimum amplification is of the order of 50, similar to the case shown on Fig. \ref{rys_1} a). Again, we note that a thinner layer yields a larger amplification factor, but it can support a smaller number of excitons. Therefore, the optical response of the thin layer particle will be weaker. On the other hand, in many cases such as in the experiments described in \cite{Neubauer}, the problem is to separate the amplified quadrupole transition from the absorption background of multiple strong dipole transitions. In such a case, limiting the number of excitons and confining them to a small volume where amplification is substantial would be beneficial.

\subsection{Cylindrical structure}
\begin{figure}[ht!]
\centering
\includegraphics[width=.95\linewidth]{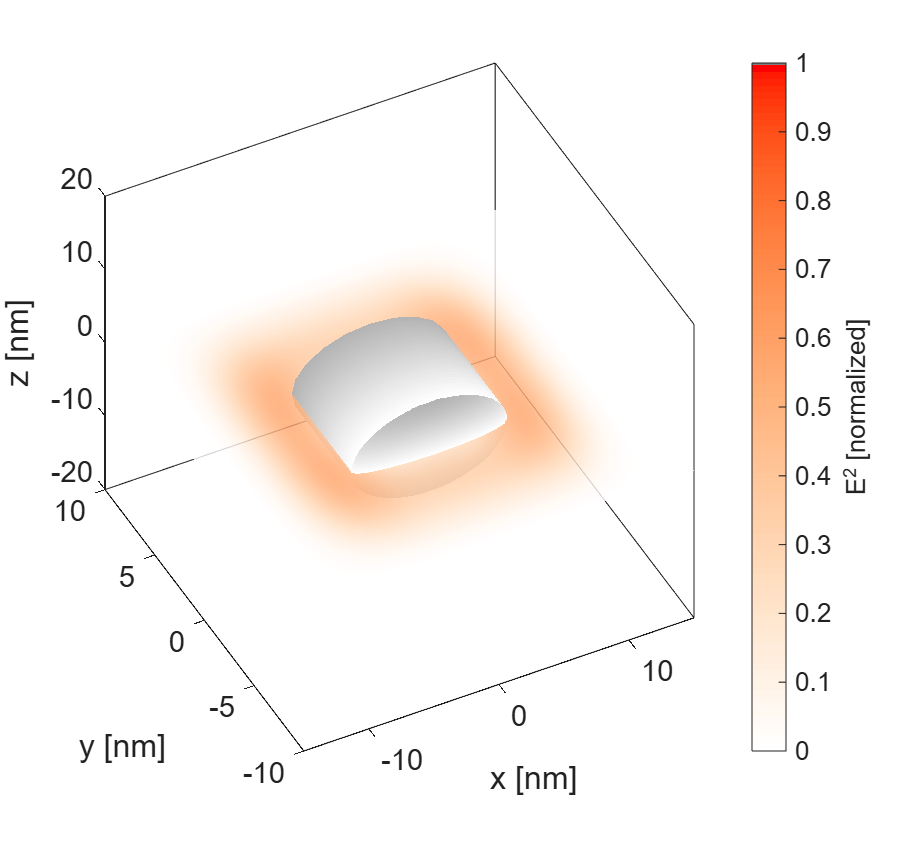}
\caption{Numerically calculated electric field intensity in the vicinity of a metal cylinder.}\label{rys_metalcyl}
\end{figure}

\begin{figure}[ht!]
\centering
a)\includegraphics[width=.8\linewidth]{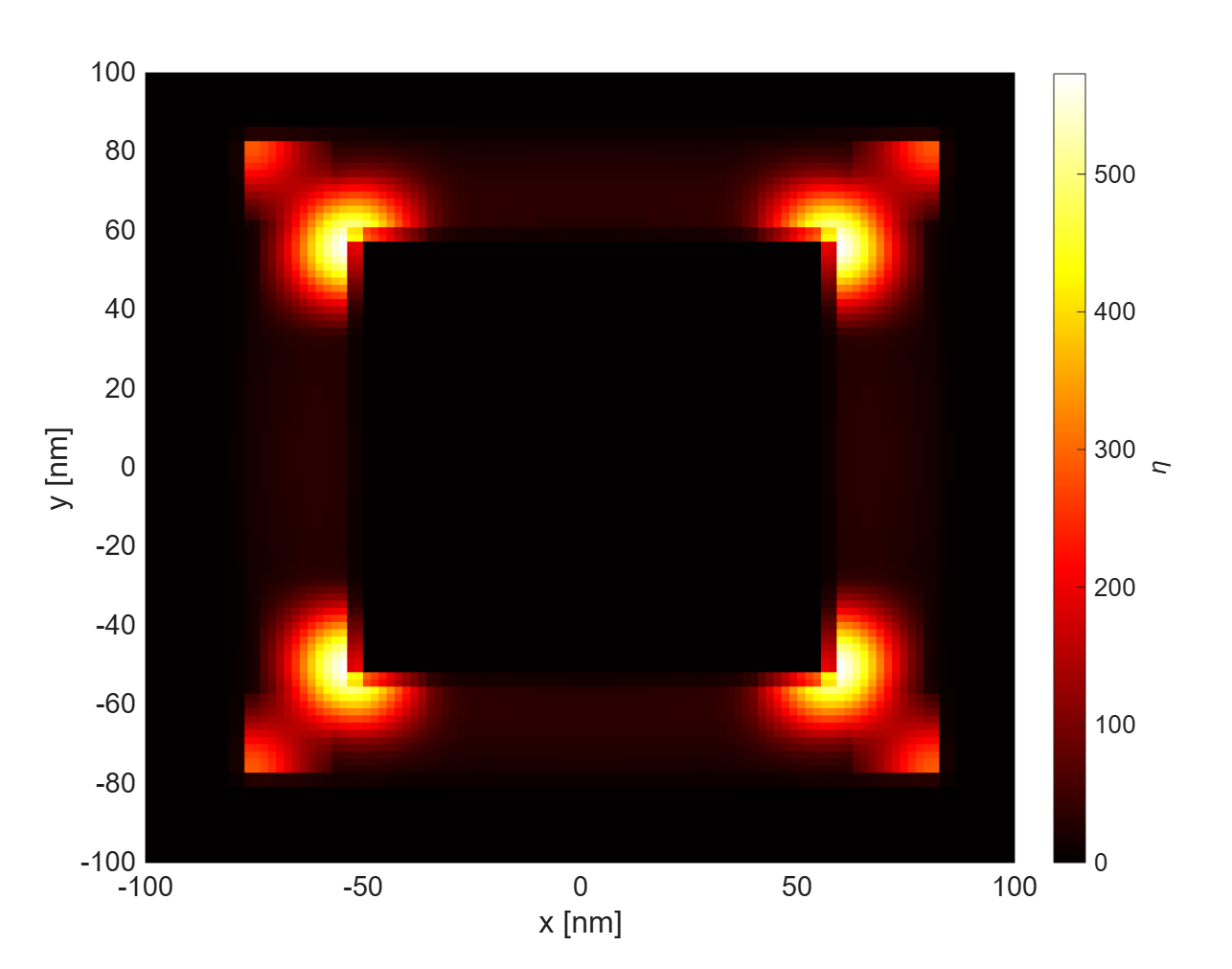}
b)\includegraphics[width=.8\linewidth]{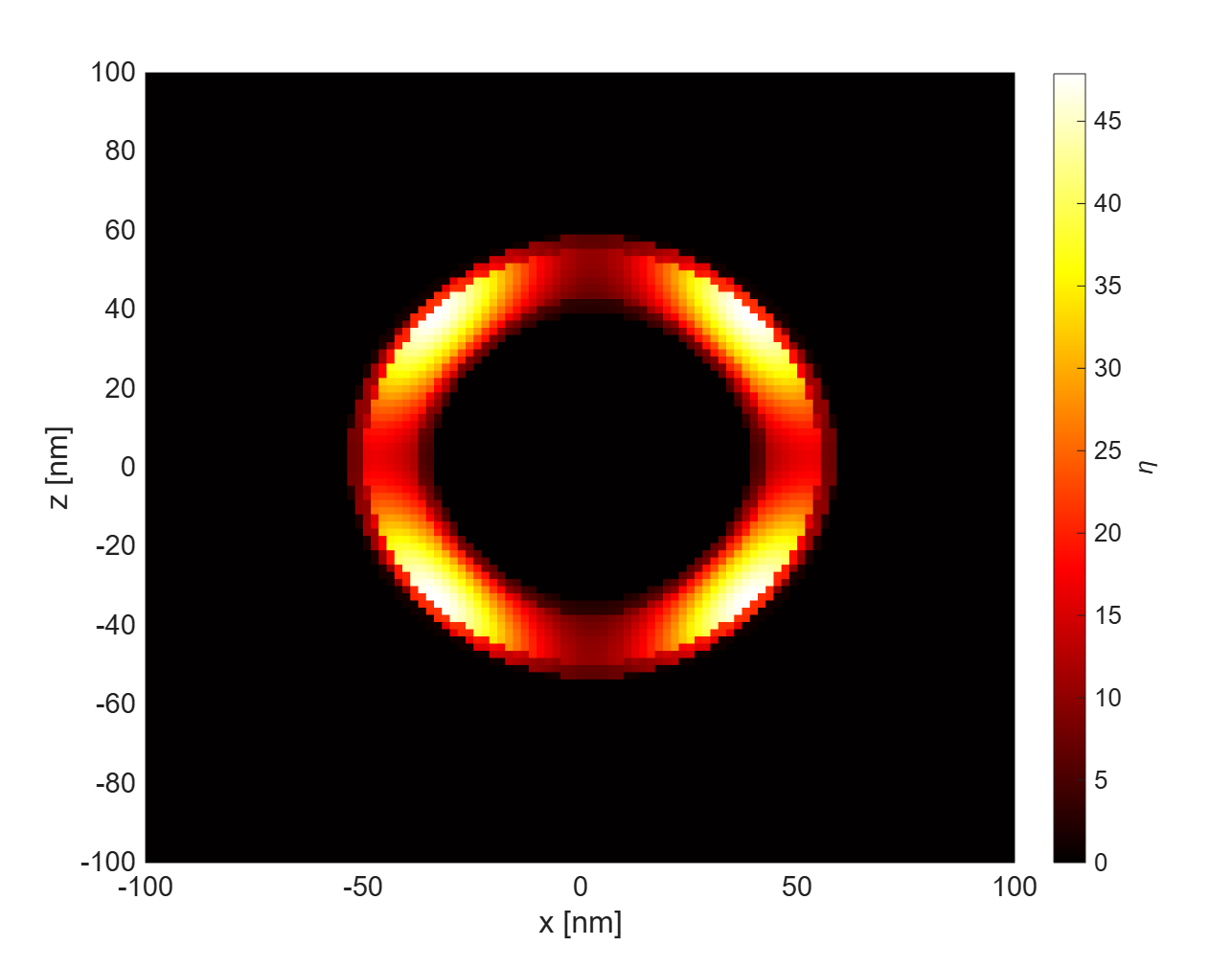}
\caption{Local amplification (color) of the quadrupole transition to 3$S$ state for a cylindrical core-shell particle. Cylinder radius is 50 nm and the height is 100 nm. A cross-section on a) $XY$ and b) $XZ$ planes are shown.}\label{rys_c1}
\end{figure}

A second type of a core-shell structure considered in this paper is a metal cylinder surrounded by Cu$_2$O layer. In such a system, multiple surface plasmon modes can be excited due to the more complicated geometry than a spherical particle. The most dominant one is a dipole mode aligned with the axis of the cylinder, as shown in Fig. \ref{rys_metalcyl}. 
One can notice that the strongest field gradients occur near the top and bottom edges of the cylinder. In these regions, the quadrupol transition amplification will be the strongest. This is the case in the results shown in Fig.  \ref{rys_c1}. The cylinder is aligned with $y$ axis. The side view on Fig. \ref{rys_c1} a) shows that the strongest field concentration (and thus quadrupole transition enhancement) is at the top and bottom edges. Like in other nanoantennas, the presence of sharp edges results in strong field gradients \cite{Okuda2006}. In these regions, the amplification factor exceeds $\eta \sim 500$. The second cross-section through the middle of the cylinder (Fig. \ref{rys_c1} b)) reveals that a quadrupole plasmonic mode has formed on the side walls. This mode is much weaker than the one on the top and bottom edges, with peak amplification of the order of $\eta \sim 50$. 

\begin{figure}[ht!]
\centering
\includegraphics[width=.9\linewidth]{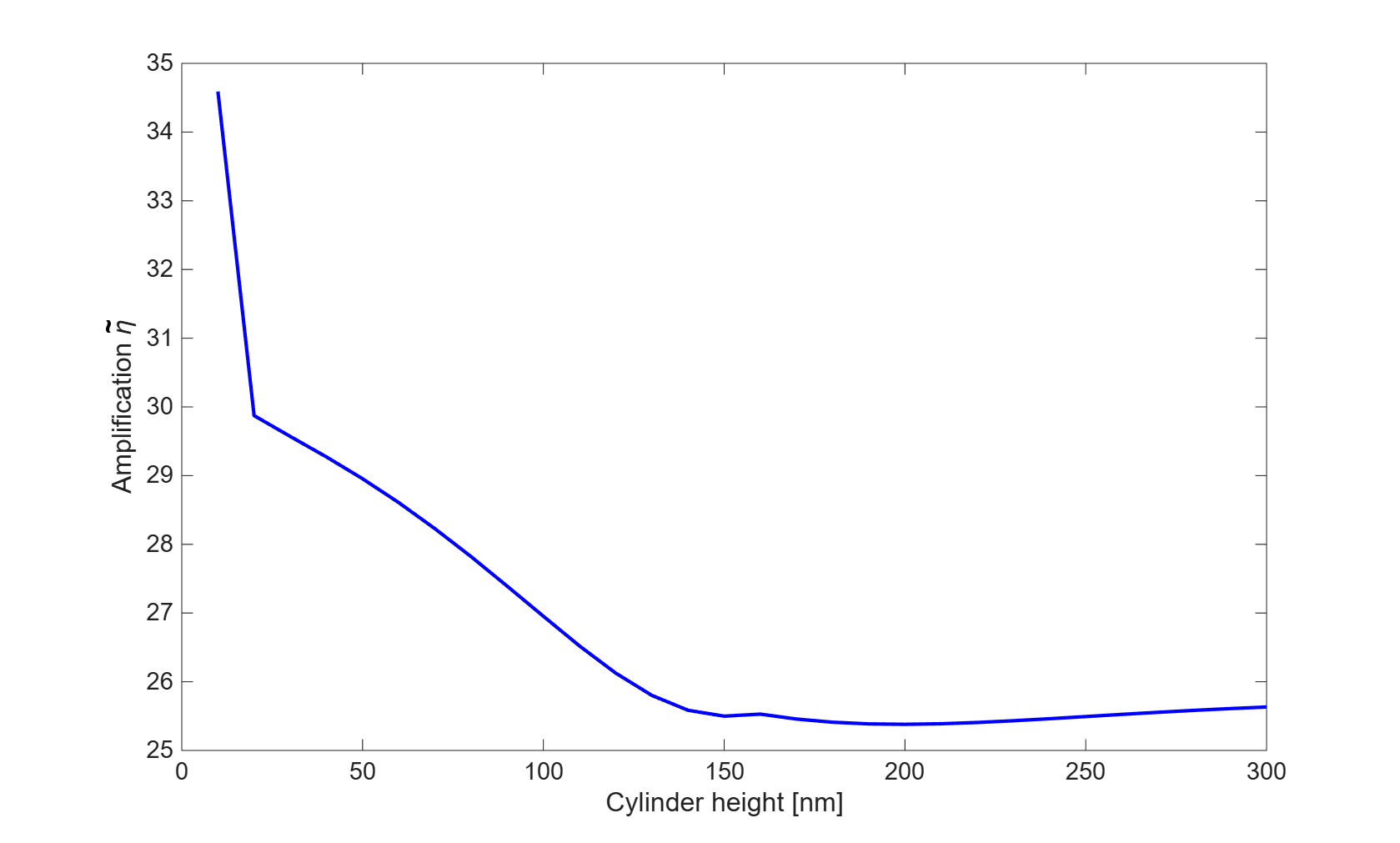}
\caption{Averaged amplification of the quadrupole transition to 3$S$ state in a cylindrical core-shell structure, as a function of cylinder height. Cylinder radius is $50$ nm, Cu$_2$O layer thickness is 30 nm.}\label{rys_c2}
\end{figure}

The above results indicate that the cylindrical core-shell particle is characterized by a strong amplification near the cylinder edges, but relatively weak amplification everywhere else. This is further illustrated in Fig. \ref{rys_c2}, where an average amplification is calculated as a function of the cylinder height.
Compared to the spherical particle results shown in Fig. \ref{rys_1b}, the maximum amplification is about 4 times smaller and occurs in the limit of a vanishing height (so that the structure becomes a nanodisc). As the height increases, the influence of the edge plasmon modes on the average amplification becomes smaller. For the height exceeding 150 nm, the amplification stabilizes at a value determined by the quadrupole plasmon mode forming on the side walls (Fig. \ref{rys_c1} b)).

\subsection{Cu$_2$O nanosphere}

\begin{figure}[ht!]
\centering
a)\includegraphics[width=.8\linewidth]{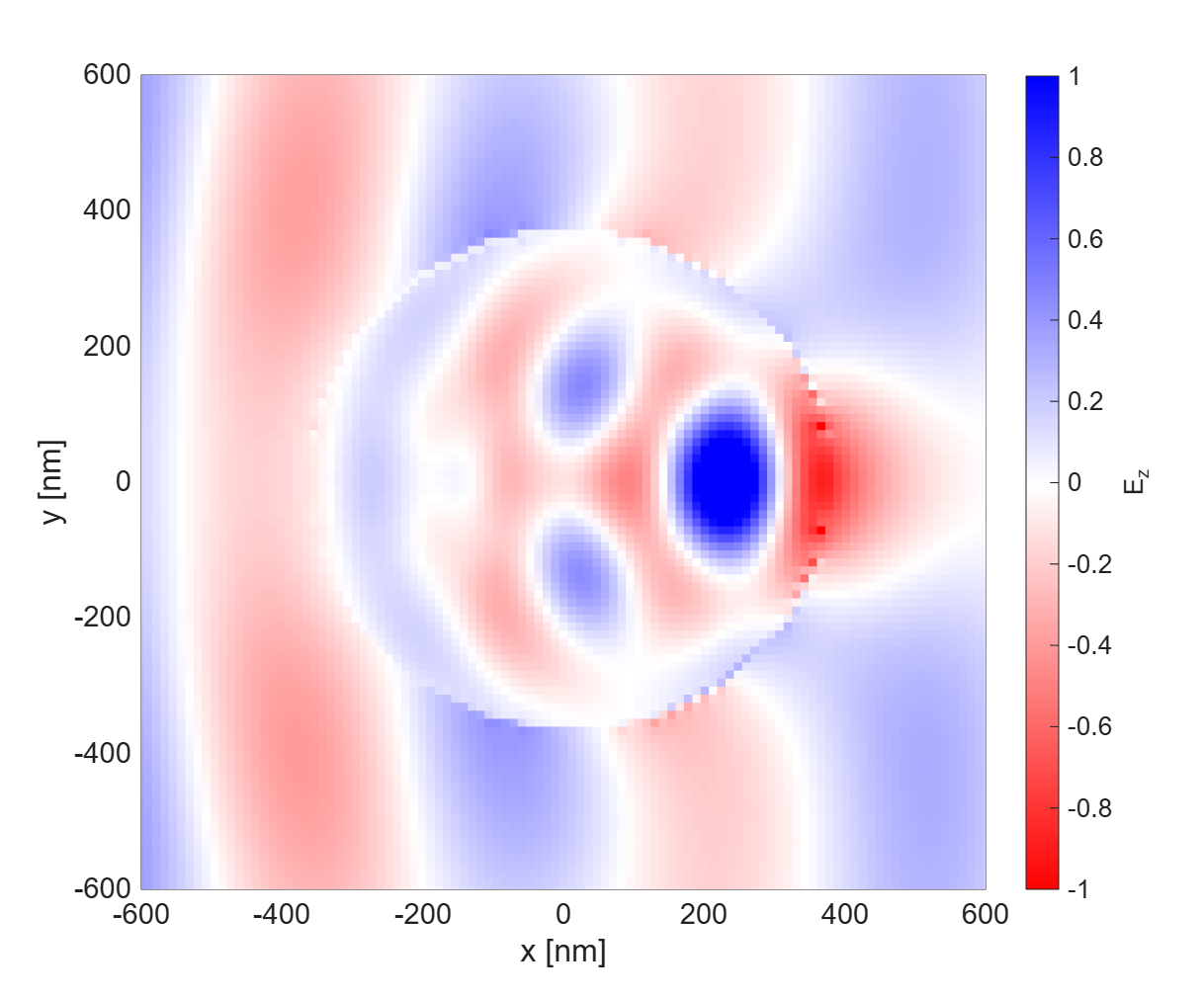}
b)\includegraphics[width=.8\linewidth]{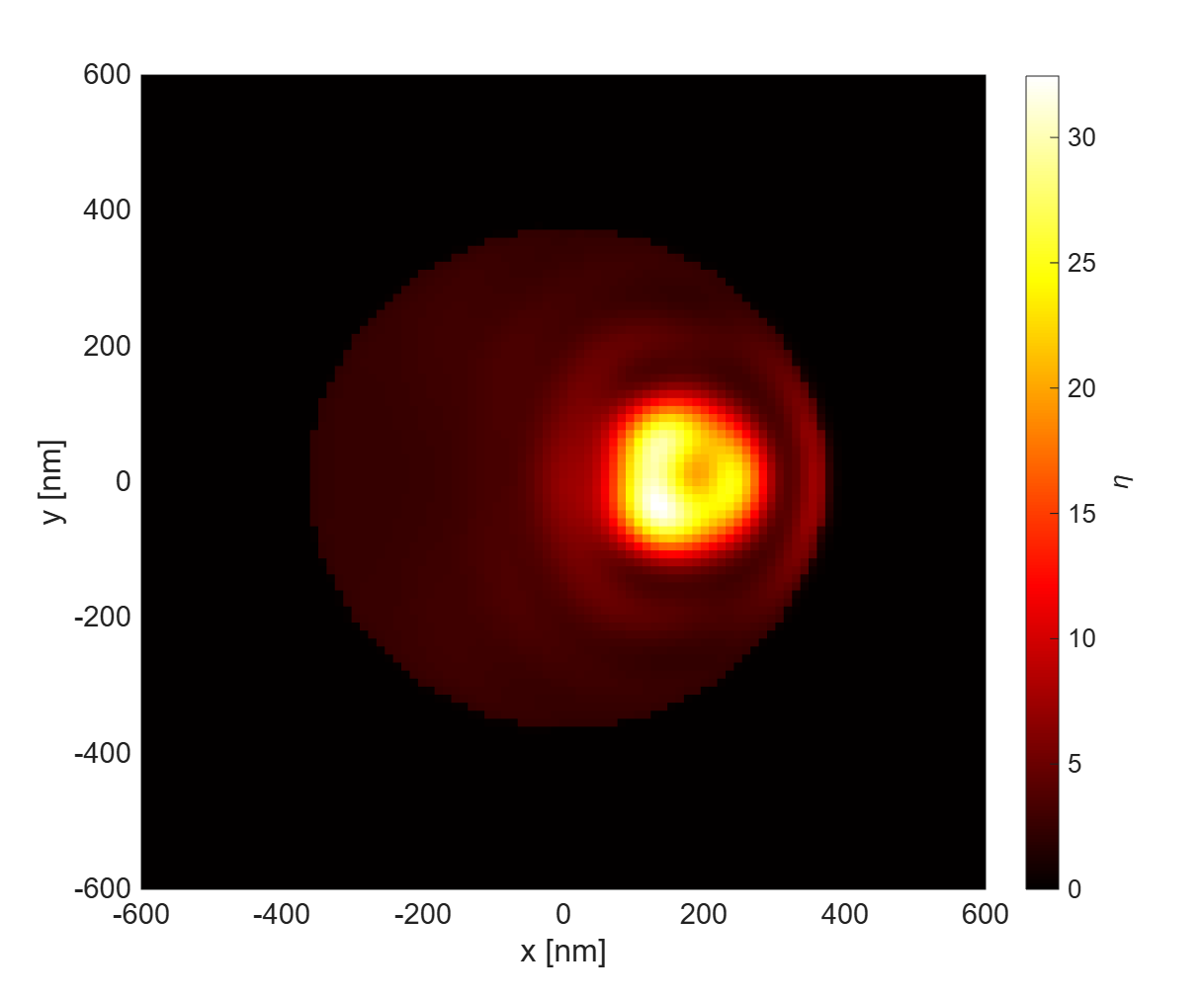}
\caption{Numerically calculated field distribution a) and quadrupole transition amplification b) of a Cu$_2$O nanosphere.}\label{rys_k1}
\end{figure}

Finally, for the sake of comparison, it is interesting to study the possible local field amplification that can be achieved even without surface plasmons. Specifically, a sphere made out of Cu$_2$O can act as a lens. The so-called ball lens has a wide range of applications in microscopy and optical fibers \cite{Schwarz2005,Kim2016}. The focal length of a ball lens is given by \cite{Kim2016}
\begin{equation}
f = \frac{n_rR}{2(n_r-1)},
\end{equation}
where $R$ is the ball radius and $n_r$ is the refraction index. Crucially, the copper oxide is characterized by a refraction index that is greater than 2, specifically $n_r \sim \sqrt{7.5} \approx 2.73$. This means that the focal point is located inside the ball. Therefore, a strong (but still diffraction limited) field concentration occurs inside Cu$_2$O. The result of simulation of such a system are shown in Fig. \ref{rys_k1}.
Fig. \ref{rys_k1} a) depicts the $x$ component of the electric field vector obtained in FDTD calculation. Outside of the sphere, plane waves are propagating along x axis, from left to right. Inside, they are focused in a focal point located at $x=200$ nm. The diameter of the sphere is 600 nm, which is comparable to the free space wavelength of incident light. In Fig. \ref{rys_k1} b), the amplification of the transition to $3S$ exciton state is presented. Here, the focal point corresponding to the strongest field gradient is more clearly visible. Its diameter is approximately 200 nm, which is comparable to the wavelength inside Cu$_2$O. Thus, unlike surface plasmons, the focusing is diffraction-limited. As a result, the peak amplification is only of the order of 30.

\section{Conclusions}
We have investigated the plasmonic amplification of the quadrupole transition between the ground state and $S$ Rydberg excitonic states in several types of Cu$_2$O nanostructures. The field averaging procedure developed in our previous work \cite{Paper1} has been adapted to take into account the large spatial extent of Rydberg excitons in Cu$_2$O. The calculation results indicate that core-shell type nanostructures with Rydberg excitons are a good candidate for strong  amplification by facilitating the excitation of multiple deep subwavelength plasmon modes. The plasmonic enhancement (i.e., an average amplification factor) strongly depends not only on the particular excitonic state $n$ but also on geometrical  properties of a given nanostructure.  Due to the match between structure size and Rydberg exciton radius, the proposed nanostructures may provide a convenient platform for single exciton manipulation. Our study follows theoretical \cite{Deguchi2009} and recent experimental \cite{Neubauer} developments, contributing to the novel field of Rydberg exciton-plasmon interaction science.

\section{Acknowledgment}
Support from National Science Centre, Poland (Project No. OPUS 2025/57/B/ST3/00334), is greatly acknowledged.

\section{Appendix A: excitonic wavefunctions in a core-shell structure}
\subsection{Spherical symmetry}

Following the method described, for example, in Ref. \cite{Landau}, we transform the Schr\"{o}dinger equation into a Whittaker equation. Like in the paper I \cite{Paper1}, we assume a confinement potential in the form
\begin{eqnarray}\label{boundarysphere}
&&V_{conf}(r_{e,h})=\left\{ \begin{array}{ll}
0\quad \mbox{for}\quad r_1\leq\quad r_{e,h}\leq r_2,\\
\infty\quad \mbox{for}\quad r_{e,h}>r_2\quad\hbox{or}\; r_{e,h}<r_1.
\end{array}\right.
\end{eqnarray}
where $r_1$ is the radius of the metal core and $r_2$ is the radius of the Cu$_2$O shell. The confinement potential is identical for the electron and hole. Due to the lower effective mass of the hole, it is more mobile; the electron is assumed to be stationary. Therefore, like in the paper I, we consider hydrogenlike wavefunctions of the hole. The Schr\"{o}dinger equation for the hole is 
\begin{eqnarray}
&&\Biggl[-\frac{\hbar^2}{2m_h}\hbox{\boldmath$\nabla$}^2\Biggr]\psi(\textbf{r}_h)=(E-V)\psi(\textbf{r}_h).
\end{eqnarray}
with
\begin{equation}
V=V_{conf}-\frac{e^2}{4\pi\epsilon_0\epsilon_b\vert r_h\vert}
\end{equation}
and
\begin{equation}
\hbox{\boldmath$\nabla$}^2=\frac{d^2}{dr_h^2}+\frac{2}{r_h}\frac{d}{dr_h}-\frac{\mathcal L^2}{r_h^2}, 
\end{equation}
where ${\mathcal L}$ is the angular momentum operator given by
\begin{equation}
{\mathcal L^2}=-\left[\frac{1}{\sin\theta}\frac{\partial}{\partial\theta}\left(\sin\theta\frac{\partial}{\partial\theta}\right)+\frac{1}{\sin^2\theta}\frac{\partial^2}{\partial\phi^2}\right].
\end{equation}
with eigenvalue
\begin{equation}
{\mathcal L^2}Y=\ell(\ell+1)Y.
\end{equation}
We assume that the solution is a hydrogenlike wavefunction $\psi_h(\textbf{r}_h)=Y_{\ell
m}(\theta,\phi) \psi(r)$, with spherical harmonics $Y_{\ell m}(\theta,\phi)$ and radial part $\psi(r_h)$.
Thus, we have
\begin{equation}
-\frac{\hbar^2}{2m_h}\left[\frac{d^2}{dr_h^2}+\frac{2}{r_h}\frac{d}{dr_h}-\frac{\mathcal L^2}{r_h^2}\right]Y\psi=(E-V)Y\psi
\end{equation}
This leads to
\begin{equation}
\frac{\partial^2 \psi}{\partial r_h^2}Y+\frac{2}{r}\frac{\partial \psi}{\partial r_h}Y-\frac{\mathcal L^2}{r_h^2}Y\psi=\frac{-2m_h}{\hbar^2}(E-V)Y\psi.
\end{equation}
The equation for radial part is
\begin{equation}
\frac{\partial^2 \psi}{\partial r_h^2}+\frac{2}{r_h}\frac{\partial \psi}{\partial r_h}-\frac{l(l+1)}{r_h^2}\psi+\frac{2m_h}{\hbar^2}(E-V)\psi=0.
\end{equation}
By using a new function $R=r_h\psi$ we get
\begin{equation}
\frac{\partial^2 R}{\partial r_h^2}-\frac{l(l+1)}{r_h^2}R+\frac{2m_h}{\hbar^2}(E-V)R=0.
\end{equation}
As mentioned above, the potential $V$ includes both the Coulomb potential $V_c$ and confinement potential $V_{conf}$. Thus, we get 
\begin{eqnarray}\label{rad_1}
&&\frac{\partial^2 R}{\partial r_h^2}+\left[\frac{2m_hE}{\hbar^2}+\frac{2m_h}{\hbar^2}\frac{e^2}{4\pi\epsilon_0\epsilon_b r_h}\right.\nonumber\\&&\left.-\frac{\ell(\ell+1)}{r_h^2}+\frac{2m_h}{\hbar^2}V_{conf}(r_h)\right]R=0.
\end{eqnarray}
Now we introduce the Rydberg radius
\begin{equation}
a_h=\frac{4\pi\epsilon_0\epsilon_b\hbar^2}{m_he^2},
\end{equation}
and the dimensionless radius 
\begin{equation}
\rho=\frac{r}{a_h},
\end{equation}
the Rydberg energy
\begin{equation}
R_h^*=\frac{\hbar^2}{2m_ha_h^2},
\end{equation}
and the scaled energy 
\begin{equation}
\frac{1}{\eta^2}=\frac{-E}{R_h^*}.
\end{equation}
Substituting to Eq. (\ref{rad_1}), one obtains
\begin{equation}\label{rad_ogolne_2}
\frac{d^2}{d\rho^2}R+\left[\frac{E}{R_h^*}+\frac{2}{\rho}-\frac{\ell(\ell+1)}{\rho^2}+\frac{V_{conf}(\rho)}{R^*}\right]R=0.
\end{equation}
Using the scaled variables 
\begin{equation}
\rho=\frac{\eta}{2}\xi, \quad \ell = \mu-\frac{1}{2},
\end{equation}
one obtains
\begin{equation}
\frac{d^2}{d\xi^2}R+\left[\frac{-1}{4}+\frac{\eta}{\xi}-\frac{\mu^2-\frac{1}{4}}{\xi^2}+\frac{\eta^2}{4}\frac{V_{conf}(\xi)}{R^*}\right]R=0.
\end{equation}
As mentioned before the confinement potential has a form of infinite potential walls, so that $V_{conf}=0$ when $r_1<r_h<r_2$ and $V_{conf} \rightarrow \infty$ otherwise. Therefore, the wavefunction in the volume inside the nanostructure fulfills the equation
\begin{equation}\label{Whitt1}
\frac{d^2R}{d\xi^2}+\left[-\frac{1}{4}+\frac{\eta}{\xi}+\frac{1/4-\mu^2}{\xi^2}\right]R=0
\end{equation}
which is a so-called Whittaker equation.

\begin{figure}[ht!]
\centering
\includegraphics[width=.9\linewidth]{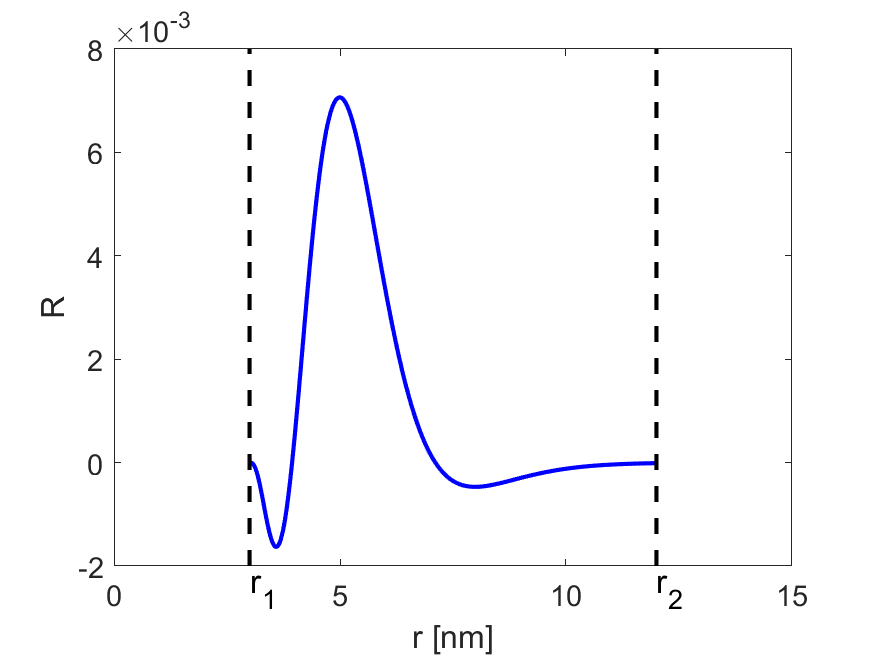}
\caption{The radial wavefunction of the hole in a core-shell structure with $r_1=3$ nm, $r_2=12$ nm, calculated for $3S$ exciton.}\label{rys_w1}
\end{figure}

As discussed in paper I \cite{Paper1}, in the intermediate case where exciton radius and the nanostructure radius are comparable, the spherical symmetry of the problem is not maintained. In order to avoid this difficulty, we have proposed a modified Coulomb potential
\begin{equation}
V_c(r)=\frac{-e^2}{4\pi\epsilon_0\epsilon_b (r-\alpha r_1)}.
\end{equation}
By putting it in the Eq. (\ref{rad_ogolne_2}), one obtains
\begin{equation}\label{rad_ogolne_3}
\frac{d^2R}{d\rho^2}+\left[\frac{E}{R_h^*}+\frac{2}{\rho-\alpha\rho_1}-\frac{\ell(\ell+1)}{\rho^2}+\frac{V_{conf}(\rho)}{R^*}\right]R=0,
\end{equation}
where $\rho_1=r_1/a_h$. Again using the scaled variables and considering the region inside the structure where $V_{conf}=0$, one obtains
\begin{equation}
\frac{d^2}{d\xi^2}R+\left[-\frac{1}{4}+\frac{\eta}{(\xi-\alpha\xi_1)}+\frac{1/4-\mu^2}{\xi^2}\right]R=0,
\end{equation}
with $\xi_1=\frac{2}{\eta}\rho_1$. By using scaling
\begin{equation}
\xi'=\frac{2}{\eta}(\rho-\alpha\rho_1)
\end{equation}
one obtains
\begin{equation}\label{Whitt_bad1}
\frac{d^2}{d\xi'^2}R+\left[-\frac{1}{4}+\frac{\eta}{\xi'}+\frac{1/4-\mu^2}{(\xi'+\xi_1)^2}\right]R=0,
\end{equation}
which is analogous to Eq. (\ref{Whitt1}) in the limit of $\xi_1 \rightarrow 0$. In the case of $S$ excitons considered here, $\mu=1/2$ and the third term in parenthesis vanishes. Then, we obtain a special case of a Whittaker equation (\ref{Whitt1}), but with scaled $\xi'$. An example solution for the radial function of $1S$-$4S$ excitons is shown in Fig. \ref{rys_w1}.

\subsection{Cylindrical symmetry}
Here, we assume that \textbf{r} is a two-dimensional vector on $xy$ plane. The Hamiltonian is 
\begin{eqnarray}\label{H_disk}
&&H=E_g+\frac{p_{ez}^2}{2m_{e}}+\frac{p_{hz}^2}{2m_{h}}\nonumber\\
&&-\frac{\hbar^2}{2m_{e}}{\hbox{\boldmath$\nabla$}_e}^{(2D)2}
-\frac{\hbar^2}{2m_h}{\hbox{\boldmath$\nabla$}_h}^{(2D)2}\nonumber\\
&&+V_e(\textbf{r}_e)+V_h(\textbf{r}_h)\nonumber\\
&&-\frac{e^2}{4\pi\epsilon_0\epsilon_b\sqrt{(\textbf{r}_e-\textbf{r}_h)^2+(z_e-z_h)^2}},
\end{eqnarray}
where $m_h$ is the hole effective masses, the electron effective mass is $m_e$,  $\epsilon_b$  is the static dielectric constant of the QR material and 
\begin{equation}
{\hbox{\boldmath$\nabla$}}^{(2D)2}=\frac{1}{r^2}\left(r\frac{\partial}{\partial
r}r\frac{\partial}{\partial
r}+\frac{\partial^2}{\partial\phi^2}\right).
\end{equation}
Again, we focus on the hole wavefunction which has the form
\begin{equation}
\Psi_h(r_h,\phi)=\psi_h(r_h)e^{i m \phi},
\end{equation}
with quantum number $m=0,1...$. The hole radial eigenfunction $\psi_h(r_h)$ satisfies the equation
\begin{eqnarray}\label{e13}
&&\left[-
\frac{\hbar^2}{2m_{h}}\frac{1}{r_h^2}\left(r_h\frac{\partial}{\partial r_h}r_h\frac{\partial}{\partial_h}-m^2\right)\right.\nonumber\\
& &\left.-\frac{e^2}{4\pi\epsilon_0\epsilon_b
r_h}+V_\parl(r_h)\right]\psi_h=E_h\psi_h. 
\end{eqnarray}
This leads to
\begin{equation}
\left[\frac{\partial^2}{\partial r_h^2}+\frac{1}{r_h}\frac{\partial}{\partial r_h}-\frac{m^2}{r_h^2}\right]\psi_h=\frac{2m_h}{\hbar^2}(E-V)\psi_h.
\end{equation}
Again, using the substitution $R=\sqrt{r_h}\psi$, one obtains
\begin{equation}
\frac{\partial^2\psi_h}{\partial r_h^2}+\frac{1}{r_h}\frac{\partial \psi_h}{\partial r_h}=\frac{1}{\sqrt{r_h}}\frac{\partial^2 R}{\partial r_h^2}+\frac{1}{4}\frac{1}{r_h^2\sqrt{r}}R
\end{equation}
and finally
\begin{equation}
\left[\frac{\partial^2}{\partial r_h^2}+\frac{1-4m^2}{4r_h^2}\right]R=\frac{2m_h}{\hbar^2}(E-V)R
\end{equation}
which is analogous to Eq. (\ref{rad_1}) but with $(1-4m^2)/4$ instead of $\ell(\ell+1)$. Thus, with scaled units we obtain
\begin{equation}\label{rad_ogolne_5}
\frac{d^2}{d\rho^2}R+\left[\frac{E}{R_h^*}+\frac{2}{\rho}+\frac{1-4m^2}{4\rho^2}+\frac{V_{conf}(\rho)}{R^*}\right]R=0.
\end{equation}
Introducing the quantities
\begin{equation}\label{e14}
\tilde{\epsilon}=-\epsilon=-\frac{E_h}{R^*_h},\quad
\xi=\alpha\rho,\quad \alpha^2=4\tilde{\epsilon},
\end{equation}
and considering the region inside the structure where $V_{conf}=0$, we obtain
\begin{equation}
\frac{d^2}{d\xi^2}R+\left[\frac{\epsilon}{\alpha^2}+\frac{2}{\xi\alpha}+\frac{1/4-m^2}{\xi^2}\right]R=0.
\end{equation}
With additional scaling $\eta=2/\alpha$ and using $\alpha^2=-4\epsilon$, one obtains
\begin{equation}
\frac{d^2R}{d\xi^2}+\left[-\frac{1}{4}+\frac{\eta}{\xi}+\frac{1/4-m^2}{\xi^2}\right]R=0,
\end{equation}
which again is a Whittaker equation.

\section{Appendix B: FDTD method}
\begin{figure}
\includegraphics[width=.75\linewidth]{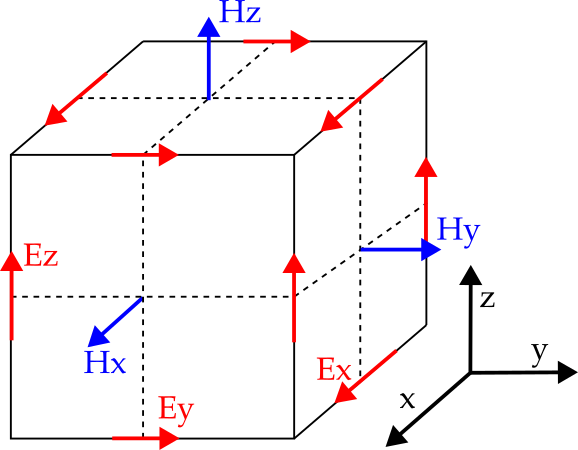}
\caption{Numerical grid used in FDTD calculations.}\label{fig:numgrid}
\end{figure}

The spatial distribution of the electric field in considered nanostructures has been calculated with the Finite-Difference Time-Domain (FDTD) algorithm \cite{Yee}. This numerical method is based directly on Maxwell's equations, providing high accuracy of predictions. In particular, we use the following field evolution equations
\begin{eqnarray}
&&\epsilon\frac{\partial E_x}{\partial t}+\sigma E_x+\frac{\partial P_x}{\partial t}=\left(\frac{\partial H_z}{\partial y}-\frac{\partial H_y}{\partial z}\right)\nonumber\\
&&\epsilon\frac{\partial E_y}{\partial t}+\sigma E_y+\frac{\partial P_y}{\partial t}=\left(\frac{\partial H_x}{\partial z}-\frac{\partial H_z}{\partial x}\right)\nonumber\\
&&\epsilon\frac{\partial E_z}{\partial t}+\sigma E_z+\frac{\partial P_z}{\partial t}=\left(\frac{\partial H_y}{\partial x}-\frac{\partial H_x}{\partial y}\right)\nonumber\\
&&\mu\frac{\partial H_x}{\partial t}+\mu\frac{\partial M_x}{\partial t}=-\left(\frac{\partial E_z}{\partial y}-\frac{\partial E_y}{\partial z}\right)\nonumber\\
&&\mu\frac{\partial H_y}{\partial t}+\mu\frac{\partial M_y}{\partial t}=-\left(\frac{\partial E_x}{\partial z}-\frac{\partial E_z}{\partial x}\right)\nonumber\\
&&\mu\frac{\partial H_z}{\partial t}+\mu\frac{\partial M_z}{\partial t}=-\left(\frac{\partial E_y}{\partial x}-\frac{\partial E_x}{\partial y}\right),
\end{eqnarray}
where individual components of electric field $\vec{E}$ and magnetic field $\vec{H}$ are calculated. In the discrete representation, the calculation domain is divided into cube unit cells with some size $\Delta x$ and the field evolution is advanced by a finite time step $\Delta t$. A schematic representation of the calculation grid is shown in Fig. \ref{fig:numgrid}. 

The final discrete equations are as follows
\begin{eqnarray}
&&\mu H_x^{t+1}(x,y+1,z+1) = \mu H_x^t(x,y+1,z+1)\nonumber\\&&+\Delta t[E_y^{t+0.5}(x,y+1,z+1)-E_y^{t+0.5}(x,y+1,z)]\nonumber\\&&-\Delta t [E_z^{t+0.5}(x,y+1,z+1)-E_z^{t+0.5}(x,y,z+1)]\nonumber\\[10pt]
&&\mu H_y^{t+1}(x+1,y,z+1) = \mu H_y^t(x+1,y,z+1)\nonumber\\&&+\Delta t[E_z^{t+0.5}(x+1,y,z+1)-E_z^{t+0.5}(x,y,z+1)]\nonumber\\&&-\Delta t [E_x^{t+0.5}(x+1,y,z+1)-E_x^{t+0.5}(x+1,y,z)]\nonumber\\[10pt]
&&\mu H_z^{t+1}(x+1,y+1,z) = \mu H_z^t(x+1,y+1,z)\nonumber\\&&+\Delta t[E_x^{t+0.5}(x+1,y+1,z)-E_x^{t+0.5}(x+1,y,z)]\nonumber\\&&-\Delta t [E_y^{t+0.5}(x+1,y+1,z)-E_y^{t+0.5}(x,y+1,z)]\nonumber\\
\end{eqnarray}
and for the E field
\begin{eqnarray}
&&(\epsilon+\sigma\Delta t/2)E_x^{t+0.5}(x+1,y,z)\nonumber\\&&=(\epsilon-\sigma\Delta t/2)E_x^{t-0.5}(x+1,y,z)-\Delta t\frac{\partial P_x^t}{\partial t}(x+1,y,z)\nonumber\\
&&+\Delta t[H_z^t(x+1,y+1,z)-H_z^t(x+1,y,z)]\nonumber\\&&-\Delta t[H_y^t(x+1,y,z+1)-H_y^t(x+1,y,z)],\nonumber\\[10pt]
&&(\epsilon+\sigma\Delta t/2)E_y^{t+0.5}(x,y+1,z)\nonumber\\&&=(\epsilon-\sigma\Delta t/2)E_y^{t-0.5}(x,y+1,z)-\Delta t\frac{\partial P_y^t}{\partial t}(x,y+1,z)\nonumber\\
&&+\Delta t[H_x^t(x,y+1,z+1)-H_x^t(x,y+1,z)]\nonumber\\&&-\Delta t[H_z^t(x+1,y+1,z)-H_z^t(x,y+1,z)],\nonumber\\[10pt]
&&(\epsilon+\sigma\Delta t/2)E_z^{t+0.5}(x,y,z+1)\nonumber\\&&=(\epsilon-\sigma\Delta t/2)E_z^{t-0.5}(x,y,z+1)-\Delta t\frac{\partial P_z^t}{\partial t}(x,y,z+1)\nonumber\\
&&+\Delta t[H_y^t(x+1,y,z+1)-H_y^t(x,y,z+1)]\nonumber\\&&-\Delta t[H_x^t(x,y+1,z+1)-H_x^t(x,y,z+1)].\nonumber\\
\end{eqnarray} 
The $E(x,y,z)^t$ is the field value at the coordinates $(x\Delta x,y\Delta y,z\Delta z)$ and time $t\Delta t$. The factor $0.5$ in the time coordinate indicates the fact that $E$ and $H$ field calculations are interleaved.

The medium polarization components $P_i, i=x,y,z$ are calculated with an additional differential equation \cite{Alsunaidi}
\begin{equation}\label{polaryzacje}
\ddot{P}+\gamma\dot{P}+\omega^2_{0} P=\frac{\omega^2_{p}}{\epsilon_\infty} E
\end{equation}
with material parameters $\gamma$, $\omega_p$, $\epsilon_\infty$ characterizing the specified optical medium. In our case, the above equation is used for the dispersive model of copper, while Cu$_2$O is characterized by a constant permittivity $\epsilon=7.5$. In the frequency domain, the above model provides $\epsilon(\omega)$ in the form 
\begin{eqnarray}\label{drude}
\vec{P}(\omega)&=&\epsilon(\omega)\vec{E}(\omega),\nonumber\\
\epsilon(\omega)&=&\epsilon_\infty+\frac{\omega_{p}^2}{\omega_{0}^2 - \omega^2 - i\gamma\omega},
\end{eqnarray} 
which reduces to a standard Drude model of metals for $\omega_0=0$.

\end{document}